# Flight Software Development for the EIRSAT-1 Mission

Maeve Doyle[a,*], Andrew Gloster[b], Conor O'Toole[b], Joseph Mangan[a], David Murphy[a], Rachel Dunwoody[a], Masoud Emam[c], Jessica Erkal[a], Joe Flanaghan[c], Gianluca Fontanesi[c], Favour Okosun[c], Rakhi Rajagopalan Nair[c], Jack Reilly[a], Lána Salmon[a], Daire Sherwin[c], Joseph Thompson[c], Sarah Walsh[a], Daithí de Faoite[c], Umair Javaid[c], Sheila McBreen[a], David McKeown[c], Derek O'Callaghan[a], William O'Connor[c], Kenneth Stanton[c], Alexei Ulyanov[a], Ronan Wall[a], Lorraine Hanlon[a]

[a] School of Physics, University College Dublin, Belfield, Dublin 4, Ireland
[b] School of Mathematics and Statistics, University College Dublin, Belfield, Dublin 4, Ireland
[c] School of Mechanical and Materials Engineering, University College Dublin, Belfield, Dublin 4, Ireland
* Correspondence: maeve.doyle.1@ucdconnect.ie

*Abstract*—**The Educational Irish Research Satellite, known as EIRSAT-1, is a student-led project to design, build, test and launch Ireland's first satellite. The on-board software for this mission is being developed using Bright Ascension's GenerationOne Flight Software Development Kit. This paper provides an overview of this kit and of EIRSAT-1's on-board software design. Drawing on the team's contrasting experience with writing entirely custom firmware for the mission's science payloads, this work discusses the impact of using a kit on the software development process. The challenges associated with the educational nature of this project are the focus of this discussion. The objective of this paper is to provide useful information for other CubeSat teams assessing software development options.**

## I. Introduction

EIRSAT-1 is a 2U CubeSat being developed at University College Dublin (UCD) as part of the Fly Your Satellite! programme run by the Education Office of the European Space Agency (ESA). The EIRSAT-1 project is an interdisciplinary, student-led effort to launch Ireland's very first satellite. The mission's primary objectives are educational, with the aim to develop the capabilities of the Irish higher education sector in space science and engineering and inspire the next generation of students towards the study of STEM subjects. To facilitate these aims, EIRSAT-1 will fly three novel experiments that have been developed at UCD [1]: the Gamma-ray Module or 'GMOD', a bespoke gamma-ray detector [2]; the ENBIO[1] Module or 'EMOD', a thermal materials experiment; and Wave-Based Control or 'WBC', a software-based attitude control experiment [3]. Custom hardware has been developed by the EIRSAT-1 team for both the GMOD and EMOD payloads. Excluding the Antenna Deployment Module, which has also been developed at UCD [4], the remainder of the CubeSat platform consists of Commercial Off The Shelf (COTS) components supplied by Clyde Space Ltd[2] [5].

EIRSAT-1's main flight software will run on a Clyde Space On-Board Computer (OBC), with the FreeRTOS real-time operating system. This OBC is a standard CubeSat kit PC-104 subsystem, that is built around a MicroSemi SmartFusion2 System on Chip (SoC), and includes an ARM Cortex M3 processor. The mission's main software, which is being written in the programming language C, is being developed using v18.3 of the Bright Ascension GenerationOne Flight Software Development Kit[3] (FSDK). In addition to the OBC-run software, custom software is also being developed for the payload microcontrollers, which are Texas Instruments MSP430 microcontrollers. The EIRSAT-1 software will undergo its first full mission test during an ambient test campaign set to begin in late 2019.

This paper provides an overview of the Bright Ascension GenerationOne FSDK and of EIRSAT-1's flight software, with a focus on how the design and development of the latter has been shaped by the FSDK. Using the team's experience with developing custom firmware for GMOD and EMOD without the use of a kit, the impact of using the FSDK on the software development process is discussed, particularly with regards to the challenges associated with a student project. Alternative software development options to the FSDK are also given and the software options followed by other CubeSat teams are considered. This work offers an insight into the EIRSAT-1 team's experience with kit-driven development, helping others to determine if an FSDK-like software solution is suited to them and their mission.

## II. Flight Software Development Approaches

The Bright Ascension GenerationOne FSDK was chosen by the EIRSAT-1 team to address the challenges associated with student-led development. In particular, the challenge and risks associated with maintaining the project's schedule (discussed further in Section V). Furthermore, as the FSDK is provided with software to interface Clyde Space COTS components, which are used on EIRSAT-1 [5], the team considered that this kit was very well suited to the project. However, due to differing mission, hardware and/or software requirements, this may not be the case for every project. Therefore, this section mentions alternative flight software development options that are available to and used by CubeSat teams.

---

1. http://www.enbio.eu/
2. https://www.aac-clyde.space/
3. https://www.brightascension.com/products/flight-software/

*A. KubOS*

KubOS[4] is an open-source software development kit and framework that builds on a customized Linux distribution. ISIS, Pumpkin and Beaglebone OBCs, as well as a selection of additional COTS devices, are all supported. Options are also provided for the programming language used for development, including for C, Python and Rust.

*B. Core Flight System (Starter Kit)*

The core Flight System (cFS) is an open-source flight software platform created by NASA's Goddard Space Flight Centre. cFS has flight heritage originating from multiple larger NASA missions, including the Lunar Reconnaissance Orbiter mission. However, more recently, the platform has been adapted to suit a wider variety of mission. The cFS Starter kit, also known as OpenSatKit[5], is a development kit that facilitates reuse of cFS software for other (including CubeSat) missions.

*C. CubedOS*

The services provided by CubedOS[6] are very similar to those provided by NASA's cFS, however this framework has been written using the SPARK/Ada programming language, and associated tools, as opposed to C. The SPARK toolset and code aim to facilitate development of a more reliable software image by removing complexity associated with other programming languages to reduce the risk of development errors.

A detailed comparison of the frameworks provided with A-C is given by [6].

*D. Open-Source CubeSat Software*

The Libre Space Foundation, in collaboration with the University of Patras, developed and constructed both software and hardware for UPSat[7], a 2U CubeSat launched in 2017. As a primarily objective of the Libre Space Foundation is to provide open-source access to space technologies, a Git repository containing UPSat's flight software is freely accessible for CubeSat teams to use.

*E. Custom*

To satisfy mission requirements, many CubeSat teams decide to develop custom flight software from the ground up (e.g. [7], [8] and [9]), without the use of software solutions like the Bright Ascension FSDK or those mentioned in this section. While discussing the impact of using Bright Ascension's FSDK on the development of EIRSAT-1's flight software in the following sections, the decision of these teams to develop software from the ground up is also considered.

III. GENERATIONONE FLIGHT SOFTWARE DEVELOPMENT KIT

The Bright Ascension FSDK facilitates rapid development of flight software for small/nano-satellite missions. This is primarily achieved using a model-based software development approach, known as Component-Based Development (CBD).

In CBD, a "component" is defined as a reusable, standalone software module that plays a specific role (e.g. logging data) and provides a set of functions and parameters related to that role. A software image with a range of functionality is then created by linking multiple of these components together in a software "deployment". Within a given project, components share a standard interface, commonly referred to as a "container", through which components interact, and which also allows for easy interchange of the components that make up a deployment. The primary aim of this approach is to separate complex software systems into simple, independent components that can be easily and individually designed, tested and maintained.

CBD relies on demands reoccurring across a type (or range) of software product(s), where reusability of components which satisfy these demands is the key concept driving the success of this approach. In the case of the FSDK, Bright Ascension have built their product around general software functions that are required for any space mission (e.g. hardware interfacing, data handling and communication protocols). Therefore, the FSDK is provided with a software framework, which provides some supplier-specific (including Clyde Space), low-level hardware interfacing and OS (including FreeRTOS) abstraction; as well as libraries of pre-validated software components, many of which have flight heritage. In addition to these components, useful development tooling (plugins for the Eclipse development platform) is also provided with the FSDK.

CubeSat missions launched with flight software that has been developed using the FSDK[8] include UKube-1, Centauri-1 and Centauri-2, Audacy Zero, SeaHawk-1 and IOD-1 GEMS.

IV. EIRSAT-1 SOFTWARE DESIGN

*A. Main Flight Software*

Grouping the main software components that are required for the EIRSAT-1 mission, the components can be divided into three distinct tiers, as shown in Fig. 1, which are:

- the Management Tier– manages when the system's functions are called;

- the Capabilities Tier– provides access to the system's functions; and

- the Hardware Tier– provides low-level interfaces to the system hardware.

These three tiers represent the different layers of abstraction in the software, where the components in the lower tiers provide abstraction for components in the upper tiers. This architecture has been very much shaped by the FSDK, where the libraries of kit-provided components have been specifically designed to allow for hardware, OS and protocol independence [10]. Use of the team's resources is also well captured in this architecture, as the development time required for different aspects of the flight software increases for components in upper tiers, compared to that of the lower tiers (i.e. as mission-specific software becomes increasingly required over standard kit-provided software).

---

4. https://www.kubos.com/kubos/
5. https://opensatkit.github.io/menu/about.html
6. http://cubesatlab.org/CubedOS.jsp
7. https://upsat.gr/
8. Information on current missions launched with software developed using the FSDK is available at https://www.brightascension.com/news-events/

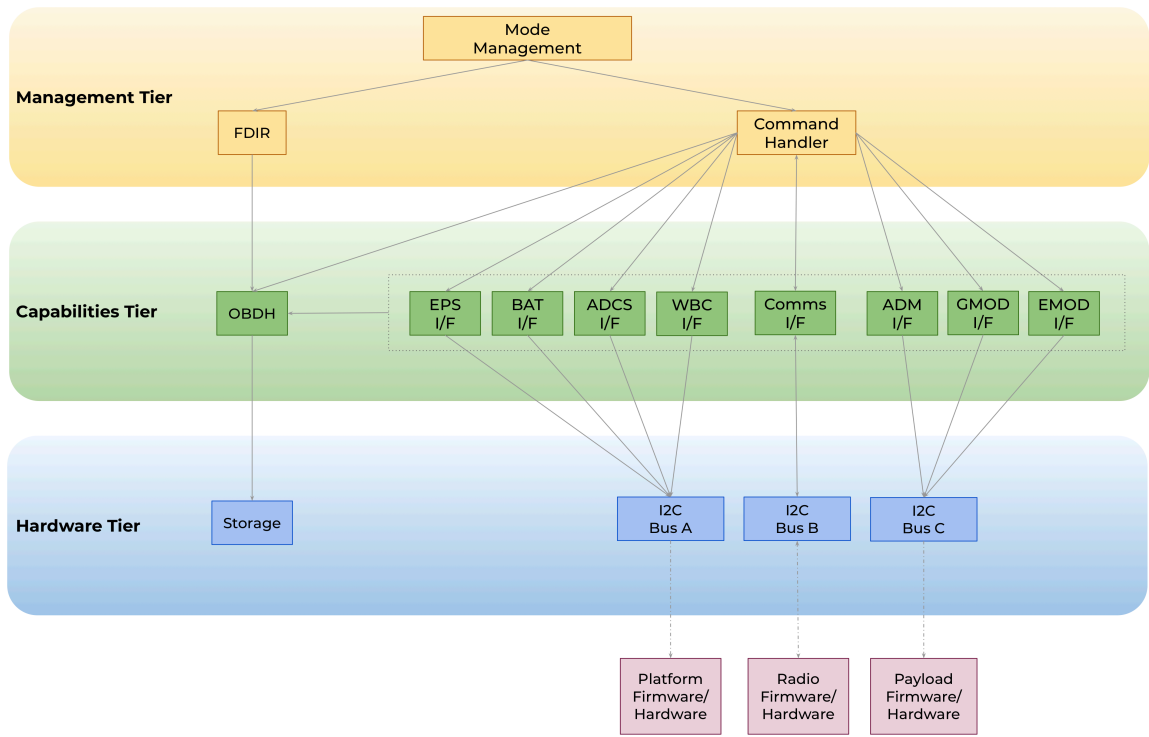

Fig. 1. Overview of the EIRSAT-1 software architecture, where *I/F* refers to the software providing an interface to the Electrical Power Supply (*EPS*), Battery (*BAT*), Attitude Determination and Control System (A*DCS*), Transciever (*Comms*), Antenna Deployment Module (*ADM*) and science experiements, *GMOD, EMOD and WBC* (uses the ADCS hardware). *FDIR* refers to the Fault Dectection, Isolation and Recovery components and *OBDH* refers to all On-Board Data Handling software.

*B. Platform and Payload Software*

Flight-ready firmware is provided by Clyde Space for the COTS platform hardware.

For the GMOD and EMOD payloads, firmware is being developed in-house without the use of development kits/tools. This firmware will provide an interface to the hardware components of the GMOD and EMOD experiments, and will communicate with the OBC via I2C. As the payload motherboards are slave devices to the OBC, the requirements on each firmware are relatively simple and are based around the need to collect and temporarily hold data in flash memory, while waiting for instruction from the OBC. As a result, the team did not adopt CBD for this development. Instead, given the limited and well-understood requirements on the systems, the payloads' firmware are being developed as a whole system in a more traditional sequential process.

Components in the main flight software then interface with this firmware to provide access to platform and payload data and functions (i.e. the interface components in Fig.1).

The architecture resulting from this development process is similar to that described for the main flight software, where software interfacing hardware is used by the software that provides functionality. However, as the payload firmware is not developed within components, little distinction between or abstraction across the tiers exists in this scenario. Furthermore, as a result of not using any development kit or tooling (i.e. developing the code from scratch), the resources of the team have been much more dedicated to lower-level aspects of the firmware as opposed to the software in the upper tiers, completely in contrast to the development time given to the different tiers of the main flight software.

V. IMPACT OF USING A KIT ON THE DEVELOPMENT PROCESS

The EIRSAT-1 team have had the opportunity to develop parts of the mission's software both with and without the aid of a development kit. Although, the latter was done for the comparatively smaller software part (i.e. the payload firmware), the different experiences had by the students working on each part allowed the team to identify how the use of a kit has impacted the flight software development process. These considerations are discussed in this section for a mission team to review when determining whether or not a development kit is suited to their project. This content specifically draws on the EIRSAT-1 team's experience with the Bright Ascension FSDK, however, many of the considerations are applicable to any development kit and have been discussed in a general context, as well as from the perspective of a student-led team.

*A. Schedule*

The FSDK was initially chosen by the EIRSAT-1 team to allow for rapid software development. As mentioned in Section III, this is primarily facilitated by providing libraries of ready-made software components as well as development tooling. However, development kits like the FSDK, are also provided with additional aids to help achieve not only fast-paced development, but also rapid learning, implementation and testing. In particular to the FSDK, this includes: well-developed learning materials, such as a detailed user manual, tutorials and example code; tools for deploying, testing and documenting software, including a unit testing framework, a mock ground segment application and methods of generating documentation directly from source code; and a long-term customer support service.

Drawing on the experiences of EIRSAT-1 team members (both new and existing), the learning and development time/effort required for the payload firmware compared to the main flight software is substantially higher. This demonstrates the effectiveness of the FSDK and its usefulness with regards to maintaining a project schedule. This is a particularly important consideration for a student-led project, where turnover of team members can be high, as (new) students complete (begin) their degrees, modules or projects through which they are involved.

*B. Flexibility*

To fully benefit from using a CBD kit with pre-existing, pre-validated components, a software project should be shaped by and built around the services that are provided with a kit (e.g. the FSDK is provided with collections of data handling, fault detection, automation and task management components [10]).

While this aspect of using a kit is very well suited to the EIRSAT-1 project (mainly due to significant flexibility of software design given the team's inexperience with flight software development) projects that are less flexible with regards to the implementation of functionality in software may not be so suited to development with a kit. In this case, development from the ground up, or using an alternative software solution (see Section II), may instead be needed to meet a mission's requirements. For example, the Galassia CubeSat, built by a student-led team at the National University of Singapore, identified that the objectives of the mission, including the objective to develop a reusable software framework for future missions, required software development from scratch [7].

*C. Risk*

The failure rate of CubeSat projects is high, with 40-50% of university-class CubeSat missions failing to achieve their primary mission objectives [11]. Furthermore, at least 2% of CubeSats are thought to fail due to software-related issues.

The FSDK is provided with libraries of robustly tested software components, many of which have flight heritage. Furthermore, this kit contains platform (for e.g. Clyde Space and Nanomind) and OS (for e.g. FreeRTOS and Linux) specific software. The EIRSAT-1 team have found these components to be highly dependable. Therefore, a high-standard kit such as the FSDK, which has been created by a team of developers with space software expertise and has in-orbit validation, can be invaluable for a project to mitigate the risk of mission failure, which is particularly prevalent for student-led projects.

*D. Cost of Resources*

Two costs should be considered: "Cost A", which is the cost of purchasing the product; and "Cost B", which is the cost of resources. Unlike Cost A, which is completely subject to whether a development kit is purchased or not, Cost B is inevitable for any software project. Cost B is the overhead cost associated with getting the software product (in this case the flight software) to a finished state. Managing Cost B, which is influenced by the resources (i.e. manpower, expertise and skills) available to a project, is essential to the success of any project. Therefore, to determine if a development kit is suited to a project, the team must first make a realistic assessment of their available resources, and then estimate the added overhead cost that would be required to develop software that fills the role of the kit-provided software, "Cost B(A)".

If a shortage of resources is identified (i.e. Cost B(A) >> Cost A) in the early stages of a project, as it was for EIRSAT-1 project (due to a lack of expertise but also due to the fact that the team is composed of masters and PhD students, many of whom have projects that are not related to the mission), and the project budget allows for it, investing in Cost A, to effectively outsource a substantial fraction of the groundwork development, is hugely advisable.

*E. Cost of Reusablilty*

A consequence of using a proprietary development kit is that no in-house intellectual property is developed with regards to a reusable software framework. In addition, the knowledge base of the team with regards mission software design is only developed within the confines of using a kit. Therefore, teams that have chosen to use a kit for their current software project, as is the case for the EIRSAT-1 team, must either: 1) re-invest in another license for the same kit or 2) overcome the costs associated with Cost B(A) for future missions. This is a major consideration for a mission team where the short-term benefits of and need for a development kit must be deliberated with the long-term plan.

In the case of EIRSAT-1, the benefits of rapid learning and development, as well as having the support of the Bright Ascension support team were considered as a necessity to the EIRSAT-1 project, particularly as this is a first-time project which must keep up with the schedule of the Fly Your Satellite! Programme. As a result, for EIRSAT-1, the need for a development kit outweighs the current need for freer future reusability of software. However, for teams with more resources (e.g. experience of developing a mission), that have a long-term plan involving multiple missions (e.g. CubeCat[9] and PolySat[10]), in-house software development becomes a more profitable consideration.

---

9. https://nanosatlab.upc.edu/en/missions-and-projects
10. http://www.polysat.org/launched and http://www.polysat.org/in-development

## VI. Discussion and Conclusions

The decision to use a development kit for a software project is hugely influenced by both the project objectives and the scenario in which the CubeSat project is being developed (e.g. development in an industry vs. academic environment). Therefore, Section V discusses the impact of using a development kit, specifically Bright Ascension's GenerationOne FSDK, on the development of EIRSAT-1's main flight software with sufficient background information on the EIRSAT-1 mission, team and software project to allow other teams to review this work and make an informed judgment on the FSDK in light of their own scenario. Although the EIRSAT-1 team do not have a comparable experience of developing a mission's main flight software without the aid of a development kit, development of payload firmware without the use of a kit or tooling was used to consider the impact of using the FSDK on the project.

This work shows that kit-driven development has been extremely beneficial to the EIRSAT-1 project, helping the team to overcome challenges faced with taking on a space software project for the first time, in an academic environment. Given this, it is interesting to note that the missions launched to-date with FSDK-developed software have primarily been commercial, as opposed to academic, student-led missions. A review of existing literature to establish what then is being used for other missions suggests that both industry and university CubeSat teams are commonly opting for full in-house development of flight software for the reasons stated within this work (i.e. development of knowledge/skills, flexibility, freer reusability, cost and available resources). Open-source materials are used by these teams to help their development. This includes using the options mentioned in Section II as reference models, but also involves using widely-used code and tools (e.g. using a Linux operating system [12, 13]), which generally relates to the level of support available from the community, as well as available documentation on coding, software design and development standards (e.g. using the IEEE Standard 1016 software design document [14]). Nevertheless, expanding on the points made in Section V, it is worth noting that CubeSat teams which opt for open-source development also face some unpredictable risks (e.g. added costs - while the general software product may be free, good quality support, documentation, bug-fixing, etc. may not). For these projects, the modular software architecture resulting from CBD is very popular for developing maintainable and extensible flight software. Other modular development approaches used by CubeSat teams include agile software development and service-oriented architecture engineering [12, 14].

## Acknowledgments

We acknowledge all students who have contributed to EIRSAT-1. The EIRSAT-1 project is carried out with the support of ESA's Education Office under the Fly Your Satellite! 2 programme. The EIRSAT-1 team would also like to thank Brian White, who has greatly supported the team in developing the mission's software. The team acknowledges support form ESA via PRODEX under contract number 4000124425. We would also like to acknowledge ESA support under contract number 4000104771/11/NL/CBi. The EIRSAT-1 team further acknowledge support from Parameter Space Ltd. MD acknowledges support from the Irish Research Council (IRC) under grant GOIP/2018/2564. CO'T, DM, LS and JT acknowledge support from the IRC under grants GOIPG/2017/1031, GOIPG/2014/453, GOIPG/2017/1525 and GOIPG/2014/684, respectively. AG acknowledges a scholarship from the UCD School of Mathematics and Statistics. JE, JR and RD acknowledge scholarships from the UCD School of Physics. SMB, JM and AU acknowledge support from Science Foundation Ireland under grant number 17/CDA/4723.

## References

[1] D. Murphy et al., "EIRSAT-1 – The Educational Irish Research Satellite", 2nd Symposium on Space Educational Activities (SSEA), Budapest, Hungary, 11-13 April 2018, SSEA-2018-73

[2] A. Ulyanov et al., "Performance of a monolithic $LaBr_3$:Ce crystal coupled to an array of silicon photomultipliers", Nuclear Instruments and Methods in Physics Research, Section A, Accelerators, Spectrometer, Detectors and Associated Equipment, 810, 107-119

[3] D. Sherwin et al., "Wave-based attitude control of EIRSAT-1, 2U CubeSat", 2nd Symposium on Space Educational Activities (SSEA), Budapest, Hungary, 11-13 April 2018, SSEA-2018-93

[4] J. Thompson et al., "Double-dipole antenna deployment system for EIRSAT-1, 2U CubeSat", 2nd Symposium on Space Educational Activities (SSEA), Budapest, Hungary, 11-13 April 2018, SSEA-2018-78

[5] Clyde Space Ltd. (CSL) and Cape Peninsula University of Technology (CPUT), COTS hardware reference documents, (ADCS) CSL ICD-25-01232 Rev. F, 2017; (Battery) CSL USM-1192 Issue C, 2016; (Comms) CPUT ICD-01-00045 Rev. B, 2017; (EPS) CSL USM-1335 Rev. A, 2016; (OBC) CSL ICD-25-025555 Rev. A, 2016; (Solar Array Side Panel) CSL ICD-25-02871 Rev. A 2017; (Structure) CSL ICD-00-04499 Rev. E, 2018

[6] O. Quiros Jimenez at al., "Development of a flight software framework for student CubeSat missions", Open Source CubeSat Workshop, 24-25 September 2018

[7] H. Askari et al., "Software Development for Galassia CubeSat – Design, Implementation and In-Orbit Validation", Joint Conference of 31st International Symposium on Space Technology and Science (ISTS), 26th International Symposium on Space Flight Dynamics (ISSFD), and 8th Nano-Satellite Symposium, (NSAT), Matsuyama, Japan, 2017

[8] S. Hishmeh, "Design of Flight Software for the KySat CubeSat Bus", IEEE Aerospace conference, 7-14 March 2009 Montana, USA, pp. 1-15

[9] J. Farkas, "CPX: Design of a Standard CubeSat Software Bus", California Polytechnic State University, 2005

[10] M. McCrum and P. Mendham, "Integrating Advanced Payload Data Processing in a Demanding CubeSat Mission", workshop at the 32nd Annual AIAA/USU Conference on Small Satellites, Logan, Utah, USA, 4-9 August 2018

[11] M. Swartwout, "Reliving 24 Years in the Next 12 Minutes: A Statistical and Personal History of University-Class Satellites", proceedings of the 32nd Annual AIAA/USU Conference on Small Satellites, Logan, Utah, USA, 4-9 August 2018, SSC18-WKVIII-03

[12] A. Lill et al., "Agile Software Development for Space Applications", Deutscher Luft- und Raumfahrkongress, 5-7 September 2017, 450309

[13] H. Leppinen, "Current Use of Linux in Spacecraft Flight Software", IEEE Aerospace and Electronic Systems Magazine, 32(10):4-13, October 2017

[14] M. Normann, "Software Design of an Onboard Computer for a Nanosatellite", masters thesis, Norwegian University of Science and Technology, 2016